\documentclass[12pt]{article}
\usepackage{latexsym,epsfig,amssymb, amsmath, cite, mcite}
\newcommand{\cN}{{\cal N}}
\newcommand{\mR}{\mathbb{R}}

\def\bfone{\relax{\rm 1\kern-.35em 1}}

\newcommand{\be}{\begin{equation}}
\newcommand{\ee}{\end{equation}}
\newcommand{\ben}{\begin{displaymath}}
\newcommand{\een}{\end{displaymath}}
\newcommand{\bea}{\begin{eqnarray}}
\newcommand{\eea}{\end{eqnarray}}

\newcommand{\bean}{\begin{eqnarray*}}
\newcommand{\eean}{\end{eqnarray*}}

\makeatletter \@addtoreset{equation}{section} \makeatother

\begin{document}

\begin{titlepage}

\begin{flushright}
\small UG-09-03 \\
\end{flushright}

\bigskip

\begin{center}

\vskip 2cm

{\LARGE \bf Gaugings at angles \\ \vskip .2cm
from orientifold reductions} \\

\vskip 1.0cm

{\bf Diederik Roest}\\

\vskip 0.5cm

{\em Centre for Theoretical Physics,\\
University of Groningen, \\
Nijenborgh 4, 9747 AG Groningen, The Netherlands\\
{\small {\tt d.roest@rug.nl}}} \\

\end{center}

\vskip 2cm

\begin{center} {\bf ABSTRACT}\\[3ex]

\begin{minipage}{13cm}
\small

We consider orientifold reductions to $\mathcal{N} = 4$ gauged
supergravity in four dimensions. A special feature of this theory is
that different factors of the gauge group can have relative angles
with respect to the electro-magnetic $SL(2)$ symmetry. These are
crucial for moduli stabilisation and De Sitter vacua. We show 
how such gaugings at angles generically arise in orientifold reductions.

\end{minipage}

\end{center}

\vspace{2cm}


\vfill

\end{titlepage}


\section{Introduction}

An important issue in string theory is the stabilisation of moduli. Compactifications to four dimensions generally lead to an abundance of scalar fields, which need to be stabilised at some point in moduli space. Flux compactifications are an attractive route to such a scenario \cite{review, *Grana, *Douglas}. In addition one would like to accomodate for a positive value of the scalar potential in this vacuum. Although at first this seemed hard to realise, there are now a number of possible models for De Sitter space-times within string theory \cite{see, *KKLT, *Burgess}.

Parallel to the `top-down' approach of string compactifications one can also take a `bottom-up' perspective. There have been systematic investigations of the possibilities for moduli stabilisation and De Sitter vacua in four-dimensional gauged supergravity, irrespective of any higher-dimensional origin. For $\mathcal{N} \geq 4$ extended supergravity, the De Sitter vacua found so far are unstable and have a value for the slow-roll parameter of order one \cite{Kallosh,Westra1,*Westra2}. For $\mathcal{N} = 2$, on the other hand, there are examples with stable De Sitter vacua \cite{Fre, *Ogetbil}. The higher-dimensional origin and relation to string theory of these cases is unknown.

In this paper we focus on $\mathcal N = 4$ supergravity, since the relevant aspects are very clear in that case. Both moduli stabilisation and De Sitter vacua crucially depend on a specific property of the gauging. First of all, the gauge group needs to be a product of factors. In addition, these gauge factors need to have different angles with respect to the electro-magnetic $SL(2)$ symmetry that rotates vectors into their electro-magnetic duals \cite{dRW}. As will be discussed  in more detail later,  without such a structure the scalar potential $V$ has an overall exponential dependence on the dilaton, making it impossible to stabilise the dilaton at a finite value of $V$. Therefore it is crucial to have a product of gauge factors with relative $SL(2)$ angles, i.e.~{\it gaugings at angles}.

Despite many results on the relation between $\mathcal N = 4$  gaugings and their higher-dimensional ancestors, see e.g.\cite{D'Auria1, *D'Auria2, AFT1,*AFT2,*DAF,SW,Prezas}, the higher-dimensional origin of non-trivial $SL(2)$ angles has never been clearly pointed out\footnote{It was anticipated  in \cite{DKPZ} that orientifold reductions involving the Romans' mass parameter and NS-NS flux would lead to non-trivial $SL(2)$ angles. However, no orientifold contributions and tadpole conditions were included (this was done subsequently for $\mathcal{N} = 1$ in \cite{VZ}). More recently, the connection between orientifold reductions of massive IIA and non-trivial $SL(2)$ angles was conjectured in \cite{Thomas}.}. In this paper we work out in detail a simple orientifold reduction and identify the resulting $\mathcal N = 4$ supergravity. The latter turns out to have gaugings at angles, thus providing a higher-dimensional origin for this feature. In particular, we show how moduli stabilisation is achieved by combining contributions to the scalar potential that originate from the bulk action and from the local source terms due to the orientifold. By clarifying the relation between gaugings at angles and orientifold reductions we aim to close the gap between the `top-down' and `bottom-up' approaches.

The organisation of this paper is as follows. In section 2, we review a number of general aspects of $\mathcal N = 4$ supergravity, after which we focus on a particular(ly useful) truncation. The structure of the gauging and scalar potential is emphasised. Section 3 discusses the orientifold reduction of IIA. Again we restrict  ourselves to the simple truncation and show the equivalence to a specific $\mathcal N = 4$ theory. Finally, section 4 contains our conclusions and a number of remarks on possible extensions and the relation to other work.


\section{$\cN = 4$ gauged supergravity}

In this section we discuss the structure of the $\cN = 4$ theory and its gaugings. We will briefly summarise the general case and focus on
a simple truncation which, while technically almost trivial, nevertheless retains the special feature of gaugings at angles that we want to highlight. In the next section this will be related to
a simple orientifold reduction of IIA. For the general $\cN=4$ discussion we follow the conventions of \cite{SW}, where more details and further references can be found.

The scalars of $D=4$, $\cN = 4$ supergravity parametrise a scalar
coset of the form
 \begin{align}
  \frac{SL(2)}{SO(2)} \times \frac{SO(6,6+n)}{SO(6) \times
  SO(6+n)} \,, \label{coset}
 \end{align}
The first factor contains the scalars of the supergravity multiplet.
It is denoted by $M_{\alpha \beta}$, for which we use the following explicit parametrisation:
 \begin{align}
  M_{\alpha \beta} = e^{\phi} \left(
   \begin{array}{cc} \chi^2 + e^{-2 \phi} & - \chi \\ - \chi & 1 \end{array} \right) \,, \qquad \alpha = (+,-) \,.
 \end{align}
The $SL(2)$ indices are raised and lowered with $\epsilon_{\alpha \beta} = \epsilon^{\alpha \beta}$, where $\epsilon^{+-} = - \epsilon^{-+} = 1$.
The second factor in \eqref{coset} is spanned by the matter
multiplets. We focus on the case of six matter multiplets,
corresponding to $n=0$. In this case it is convenient to use
light-cone coordinates for the $SO(6,6)$ group. The invariant metric is of the
form
 \begin{align}
  \eta_{MN} = \eta^{MN} = \left(
   \begin{array}{cc} & \mathbb{I}_6 \\ \mathbb{I}_6 & \end{array} \right) \,, \qquad M=(1,\ldots,6,\bar
1,\ldots, \bar 6) \,. \label{eta}
 \end{align}
The corresponding $SO(6,6)$ element that parametrises the scalar
coset is denoted by $M_{MN}$. We will introduce an explicit
parametrisation later. Together with the Einstein-Hilbert term for
the metric, the scalars have the following kinetic terms\footnote{We have multiplied the total action of \cite{SW} with a factor of two.}:
 \begin{align}
  {\cal L}_{\rm kin} = \sqrt{-g} [ R + \tfrac14 \partial_\mu M_{\alpha \beta}
  \partial^\mu M^{\alpha \beta} + \tfrac{1}{8} \partial_\mu M_{MN}
  \partial^\mu M^{MN} ] \,. \label{kin1}
 \end{align}

In addition the theory contains $12+n$ vectors, transforming in the fundamental representation of $SO(6,6+n)$. A noteworthy feature is that under
the compact part of the $SL(2)$ symmetry these transform into their electro-magnetic dual. This symmetry is therefore
only realised on-shell. This is a particular feature of four-dimensional theories and leads to the following intricate structure of gaugings.

The possible gaugings of this theory have been classified within the
framework of the embedding tensor \cite{ET1,*ET2}. It turns out that one can
introduce two $SO(6,6)$ representations of gauge
parameters: an anti-symmetric three-form $f_{\alpha MNP}$ and a
fundamental $\xi_{\alpha M}$, both of which transform as a doublet
under $SL(2)$.
Consistency of such gaugings requires a number of quadratic constraints on the embedding tensor, which can be seen as
generalised Jacobi identities. For later purposes we will give the constraints for the case with $\xi_{\alpha M} = 0$, for which one finds
 \begin{align}
  f_{\alpha R[MN} f_{\beta PQ]}{}^R = 0 \,, \quad
  \epsilon^{\alpha \beta} f_{\alpha MNR} f_{\beta PQ}{}^R = 0 \,.
  \label{QC}
 \end{align}
The combination of supersymmetry and gaugings induce the following scalar potential:
 \begin{align}
  {\cal L}_{\rm pot} = & - \sqrt{-g} V \,, \notag \\
  V= & \tfrac{1}{8} f_{\alpha MNP} f_{\beta QRS} M^{\alpha
  \beta} [ \tfrac13 M^{MQ} M^{NR} M^{PS} + (\tfrac23 \eta^{MQ} - M^{MQ} )\eta^{NR}
  \eta^{PS} ] + \notag \\
  & - \tfrac{1}{18} f_{\alpha MNP} f_{\beta QRS} \epsilon^{\alpha
  \beta} M^{MNPQRS} +\tfrac{3}{8} \xi_\alpha^M \xi_\beta^N
  M^{\alpha \beta} M_{MN} \,, \label{V}
 \end{align}
where the definition of $M^{MNPQRS}$ in terms of $M^{MN}$ can be found in \cite{SW}.

As mentioned before, an important aspect of this four-dimensional supergravity is that vectors are transformed into their electro-magnetic dual under the $SO(2) \subset SL(2)$ symmetry. This on-shell symmetry is responsible for the $SL(2)$ doublet structure of the gauge parameters. Depending on the $SL(2)$ orientation, the embedding tensor picks out a vector or its dual (or a linear combination) to gauge a part of the global symmetry of the theory. Moreover, when the gauge group is a product of different factors, it is possible to choose a different $SL(2)$ orientation for the different factors. In terms of the embedding tensor, this corresponds to
 \begin{align}
  f_{\alpha MNP} = \sum_i \delta_\alpha^{(i)} f_{MNP}^{(i)} \,, \quad \xi_{\alpha M} = \sum_i \delta_\alpha^{(i)}   \xi_{M}^{(i)} \,,
 \end{align}
where $f_{MNP}^{(i)}$ and $\xi_{M}^{(i)}$ specify a factor of the gauge group and the $\delta_\alpha^{(i)}$ do not necessarily point in the same $SL(2)$ directions. This possibility is referred to as different $SL(2)$ (or $SU(1,1)$) or De Roo-Wagemans angles \cite{dRW}. If all the $SL(2)$ factors are identical, one can always rotate these to the $\alpha = +$ direction, corresponding to a zero angle. It follows from \eqref{V} that in such cases the scalar potential has an overall dependence of $e^{\phi}$ and hence a runaway direction. Therefore gaugings at angles play a crucial role in moduli stabilisation \cite{Westra1, *Westra2}.

Instead of the full $\cN = 4$ supergravity we consider the following truncation. The $SO(6,6)$ symmetry can be decomposed into
 \begin{align}
  SL(3) \times SL(3) \times \mR^+ \times
  \mR^+ \;\; \subset \;\; SL(6) \times \mR^+ \;\; \subset \;\; SO(6,6) \,. \label{SL}
 \end{align}
We focus on the subsector of the theory that is invariant under both $SL(3)$ factors. The group-theoretic nature of this truncation guarantees its consistency.

A drastic consequence is that all vectors are projected out. This follows from the decomposition of the
fundamental representation of $SO(6,6)$ into $SL(3) \times SL(3)$
(omitting the $\mR^+$ weights):
 \begin{align}
  {\bf 12} \rightarrow ({\bf 3}, {\bf 1}) \oplus ({\bf 3'}, {\bf 1}) \oplus ({\bf 1}, {\bf 3}) \oplus
  ({\bf 1}, {\bf 3'}) \,,
 \end{align}
where no singlets appear.

In the scalar sector, the $SL(2)$ scalars are unaffected by this
truncation. In
contrast, from the decomposition of the adjoint representation one
learns that many of the $SO(6,6)$ scalars are projected
out:
 \begin{align}
  {\bf 66} \rightarrow ({\bf 1},{\bf 1}) \oplus ({\bf 1} \oplus {\bf 3} \oplus {\bf 3'}, {\bf 1} \oplus {\bf 3} \oplus {\bf 3'})
   \oplus ({\bf 1}, {\bf 8}) \oplus ({\bf 8},{\bf 1}) \,.
 \end{align}
Since there are only two singlets, the truncation preserves two dilatonic scalars.
One can take the following parametrisation of the $SO(6,6)$ element $M_{MN}$ in terms of these scalars $\varphi_1$ and $\varphi_2$:
 \begin{align}
  M_{MN} = \left(
   \begin{array}{cccc} e^{-\sqrt{2/3} \; \varphi_1} &  &  &  \\
     & e^{-\sqrt{2/3} \; \varphi_2} &  &  \\
     &  &  e^{\sqrt{2/3} \; \varphi_1} &  \\
     &  &  & e^{\sqrt{2/3} \; \varphi_2}
   \end{array} \right) \otimes \mathbb{I}_3 \,.
 \end{align}
Inserting this in \eqref{kin1} gives rise to the following kinetic terms for the four scalars
that survive the truncation:
 \begin{align}
  {\cal L}_{\rm kin} = \sqrt{-g} [ R - \tfrac12 (\partial \phi)^2 -
  \tfrac12 e^{2\phi} (\partial \chi)^2 - \tfrac12 (\partial
  \varphi_1)^2 - \tfrac12 (\partial \varphi_2)^2 ] \,, \label{kin2}
 \end{align}
where we have also included the Einstein-Hilbert term for the
metric.

We now come to effect of the truncation of the embedding tensor.
As the components $\xi_{\alpha M}$ also transform in the fundamental
representation of $SO(6,6)$, these suffer the same fate as the vectors, and are
all projected out. The other components $f_{\alpha MNP}$ give rise
to a number of $SL(3) \times SL(3)$ representations, including four singlets (omitting the $SL(2)$ doublet structure):
 \begin{align}
  {\bf 220} \rightarrow 4 \cdot ({\bf 1},{\bf 1}) \oplus {\text{non-singlet representations}} \,,
 \end{align}
So there are four $SO(6,6)$ components that survive the truncation. In our light-cone basis these
correspond to $f_{\alpha 123}$, $f_{\alpha 456}$, $f_{\alpha \bar 1
\bar 2 \bar 3}$ and $f_{\alpha \bar 4 \bar 5 \bar 6}$. Moreover, the
quadratic constraints \eqref{QC} result in the simple
conditions
 \begin{align}
  f_{\alpha 123} f_{\beta \bar 1 \bar 2 \bar 3} = 0 \,, \qquad
  f_{\alpha 456} f_{\beta \bar 4 \bar 5 \bar 6} = 0 \,.
 \end{align}
Hence there are four possibilities of gauge parameters in this
truncated theory, taking either non-zero `unbarred' or `barred'
components in the $(123)$ and independently in the $(456)$
directions. 

These four models can in fact be related to each other
by particular elements of the global symmetry, which interchange the two types of light-cone directions. For example, an $SO(6,6)$ transformation of the form \eqref{eta} interchanges the six `unbarred' and `barred' directions. For an odd number of interchanged directions this transformation needs to be accompanied by 
a sign flip of the $SL(2)$ axion. Therefore the four models are physically equivalent, and in the following we will only 
consider the case with gauge parameters $f_{\alpha 123}$ and $f_{\beta 456}$.

Note that the model has a product of gauge factors\footnote{This is a slight abuse of notation, as the truncated model does not have any vectors. However, by including the fields that have been truncated out, this model can be restored to a unique $\mathcal N = 4$  supergravity with a gauging defined by these parameters.}: one in the $(123)$ directions and one in the $(456)$ directions. These gaugings are specified by four real parameters: two can be seen as gauge coupling constants while the other two correspond to the $SL(2)$ angles of the two gauge factors. One of the angles can be set equal to zero, i.e.~point in the $\alpha = +$ direction, by an $SO(2) \subset SL(2)$ transformation. We will use this to rotate away $f_{- 123}$. Moreover, if $f_{- 456}$ does not vanish, one can perform an $SL(2)$ transformation that shifts the axion to set the second angle to 90 degrees. This corresponds to setting $f_{+ 456} = 0$. In the case of two different angles, one can therefore always take these orthogonal. We will not use this and keep the second angle arbitrary, however.

Let us analyse the form of the scalar potential and its extrema for the truncated model. By writing out the general scalar potential \eqref{V} and using $f_{-123} = 0$, one finds:
   \begin{align}
    V = & \tfrac14 \big( f_{+ 123} e^{\phi/2 + \sqrt{3/2} \,
    \varphi_1} - f_{- 456} e^{-\phi/2 + \sqrt{3/2} \,
    \varphi_2} \big)^2 +  \notag \\
   & + \tfrac14 \big( f_{+ 456} + \chi f_{-
    456} \big)^2 e^{\phi + \sqrt{6} \, \varphi_2} \,. \label{pot}
   \end{align}
Strikingly, the potential combines into the sum of two squares and is positive definite. This relies crucially on the different $SL(2)$ angles: the crossterm in the first square is independent of the $SL(2)$ scalars and comes from the last line of \eqref{V}. Only in the presence of such terms can one have moduli stabilisation. In the extremum with respect to $\chi$, the term on the second line vanishes. The remaining square has an
extremum if and only if $f_{+ 123}$ and $f_{- 456}$ have equal signs, while the dilatonic scalars are such that the first square in the potential vanishes as well.
In the extremum both squares that make up the scalar potential
vanish, and we have a Minkowski solution.

Next, we investigate the issue of stability. The axion can be seen to decouple from this issue as $\delta_\chi \delta_\chi V$ is
positive while $\delta_\chi \delta_{\vec \phi} V$ vanishes, where $\vec \phi$ represents the three dilatonic scalars and $\delta_\chi
\equiv \delta / \delta \chi$ etc. The matrix $\delta_{\vec \phi} \delta_{\vec \phi} V$ turns out to have one positive and two
vanishing eigenvalues. The Minkowski solution is therefore a minimum of the scalar potential - at least in the 
truncation to $SL(3) \times SL(3)$ invariant scalars that we consider.

An interesting question is which gaugings are induced by gauge
parameters of the form above. The answer can be found in \cite{CSO},
where so-called $CSO(p,q,r)$-gaugings are considered. These groups
can be seen as group contractions of $SO(p',q')$ with $p'+q'=p+q+r$.
It turns out that each component of the structure constants that we
consider induces a $CSO(1,0,3)$ gauging\footnote{Modulo two typo's in these expressions, the
$CSO(1,0,3)$ structure constants given in appendix B of \cite{CSO}
in a Cartesian basis correspond to $f_{123}$ in our
light-cone basis.} inside $SO(6,6)$. Our gauge
group therefore consists of a product of two such factors. The total
dimension of these gauge groups is twelve, in accordance with the
number of vectors. 
Reference \cite{CSO} also performed a stability analysis with respect to
all scalars and found a number of unstable directions. The Minkowski
solution is therefore a saddlepoint of the full $\cN = 4$ theory. 

\section{Orientifold reduction of IIA}

In this section we will consider a simple orientifold reduction of the IIA theory, which will be related to the previous $\mathcal N = 4$ truncation. Further details on different aspects and more complicated cases can be found in e.g.~\cite{D'Auria1, *D'Auria2, AFT1,*AFT2, *DAF, DKPZ, VZ, DeWolfe, *Camara}.

Consider the toroidal reduction of massive IIA to four dimensions. Introducing an O6-plane corresponds to modding out by $(-)^{F_L} \, \Omega \, I_{7,8,9}$. Here $(-)^{F_L}$ and $\Omega$ correspond to the left-moving fermion number and world-sheet parity, respectively, whose combined action on the IIA bosonic fields is
 \begin{align}
  \{ \hat{g}_{\mu \nu}, \hat \phi, \hat C_3, \hat C_7  \} \quad & \rightarrow \quad + \{ \hat{g}_{\mu \nu}, \hat \phi, \hat C_3, \hat C_7  \} \,, \notag \\
  \{ \hat B, \hat C_1, \hat C_5, \hat C_9 \} \quad & \rightarrow \quad - \{ \hat B, \hat C_1, \hat C_5, \hat C_9 \} \,. \label{fields}
 \end{align}
In addition, the space-time parity operation $I_{7,8,9}$ reverses the sign of three of the coordinates on the torus:
 \begin{align}
  \{ x^7, x^8, x^9 \} \quad \rightarrow \quad - \{ x^7, x^8, x^9 \} \,. \label{coords}
 \end{align}
The indices in \eqref{fields} are taken inside the O6-plane, i.e.~in the directions $(0,\ldots,6)$. Other components with indices transverse to the O6-plane will acquire additional signs due to \eqref{coords}. Furthermore the Romans' mass parameter $\hat G_0$ of IIA is invariant under the above involution.

Instead of the general orientifold reduction we will focus on the following truncation. Consider the two $T^3$'s in the directions $\{ x^4, x^5, x^6 \}$ and $\{  x^7, x^8, x^9 \}$. Diffeomorphisms leaving the two factors separately invariant generate an $SL(3) \times SL(3)$ symmetry in the four-dimensional
description. Completely analogous to the truncation of the $\cN = 4$ theory of the previous section, we will retain only singlets with respect to both factors.

The most general  Ansatz for the ten-dimensional metric that is consistent with $SL(3) \times SL(3)$ invariance is of the form
 \begin{align}
  \hat g_{\hat \mu \hat \nu} = \left( \begin{array}{ccc} e^{\sqrt{3}/2 \, \sigma_2} g_{\mu \nu} & & \\
  & e^{-\sigma_2 / 2 \sqrt{3} + \sigma_3 / \sqrt{3}} \mathbb{I}_3 & \\
  & & e^{-\sigma_2 / 2 \sqrt{3} - \sigma_3 / \sqrt{3}} \mathbb{I}_3 \end{array} \right) \,,
 \end{align}
consisting of the four-dimensional metric $g_{\mu \nu}$ and two scalars $\sigma_2$ and $\sigma_3$. Both the Kaluza-Klein vectors and other scalars, parametrising deformations of the internal torus, are projected out by the truncation to $SL(3) \times SL(3)$ invariant fields. Furthermore, the normalisation of the $\sigma$'s is chosen to ensure canonical normalisation. The ten-dimensional bulk action is\footnote{Our IIA conventions agree with e.g.~\cite{Polchinski}. To avoid cluttering our formulae we have set $4 \pi^2 \alpha' = 1$.}
 \begin{align}
 \hat S & = 2 \pi \int d^{10}x (\hat {\mathcal L}_1 + \hat {\mathcal L}_2) \,,
 \end{align}
where the first term contains the Einstein-Hilbert term and the dilaton kinetic term, while the second term is concerned with the gauge potentials. For the first part, after reduction to four dimensions we find
 \begin{align}
 \hat {\mathcal L}_1 & = \sqrt{- \hat g} \, [ \hat R - \tfrac12 (\partial \hat \phi)^2 ] \quad \rightarrow \quad
  {\mathcal L}_1  = \sqrt{- g} \, [ R - \tfrac12 \sum_{i=1,2,3} (\partial \sigma_i )^2 ] \,. \label{L1}
 \end{align}
where we have set $\hat \phi = \sigma_1$. Note that we use Einstein frame both in ten and in four dimensions.

Next, we turn to the gauge potentials. The NS-NS two-form potential is odd under \eqref{fields} and hence has to wrap an odd cycle in the torus. However, there is only one such form that is invariant under $SL(3) \times SL(3)$: a three-form. The field strength of this gauge potential therefore only gives a constant parameter $h_3$:
 \begin{align}
  \hat H = d \hat B = h_3 \, dx^7 \wedge dx^8 \wedge dx^9 \,.
 \end{align}
The R-R gauge potentials are either even or odd. First of all, the Romans' mass parameter, which can be seen as a zero-form R-R field strength, is even and also gives rise to a constant parameter in four dimensions: $\hat{G_0} = g_0$. The R-R vector is odd under the orientifold involution. Its field strength necessarily vanishes,
 \begin{align}
  \hat G_2 = d \hat C_1 + \hat G_0 \hat B = 0 \,,
 \end{align}
as there are no odd $SL(3) \times SL(3)$ invariant zero-, one- or two-cycles on the torus. Finally, the R-R three-form is even. Its magnetic part will be proportional to the even $SL(3) \times SL(3)$ invariant three-form and give rise to a scalar $\chi$,
 \begin{align}
  \hat G_4^{\rm (m)} = d \hat C_3 - \hat H \wedge \hat C_1 + \tfrac12 \hat B \wedge B = d \chi \wedge dx^4 \wedge dx^5 \wedge dx^6 \,.
 \end{align}
It can also have an electric part. This will be more conveniently described in terms of the dual field strength, which is related by $\hat G_6^{\rm (m)} = e^{\hat \phi /2} \star \hat G_4^{\rm (e)}$. The dual five-form gauge potential is odd under \eqref{fields} and can wrap the total six-torus:
 \begin{align}
  \hat G_6^{\rm (m)} = d \hat C_5  - \hat H \wedge \hat C_3 + \tfrac16 \hat B \wedge B \wedge B = (g_6 + h_3 \chi) \, dx^4 \wedge \cdots \wedge dx^9 \,.
 \end{align}
Quantisation of these parameters requires $g_0$, $h_3$ and $g_6$ all to be integer. Moreover, we will assume $g_0 h_3$ to be positive, for reasons that will become clear later.

With the Ans\"{a}tze above, the kinetic terms for the IIA gauge potentials
reduce to a kinetic term for $\chi$ and potential terms for the three constants $h_3$, $g_0$ and $g_6$:
 \begin{align}
  \hat {\mathcal L}_2 =  \sqrt{- \hat g} \, [ & - \tfrac{1}{2} e^{- \hat \phi} \hat H \cdot \hat H - \tfrac{1}{2}  e^{5/2 \hat \phi} \hat{G_0}{}^2  - \tfrac{1}{2} e^{\hat \phi /2} \hat G_4^{\rm (m)} \cdot \hat G_4^{\rm (m)} - \tfrac{1}{2}  e^{-\hat \phi /2} \hat G_6^{\rm (m)} \cdot \hat G_6^{\rm (m)} ] \quad \rightarrow \notag \\
  {\mathcal L}_2 = \sqrt{- g} \, [ & - \tfrac{1}{2} e^{\sigma_1/2 + \sqrt{3}/2 \, \sigma_2 - \sqrt{3} \, \sigma_3} (\partial \chi)^2
  - \tfrac{1}{2}  h_3{}^2 e^{- \sigma_1 + \sqrt{3} (\sigma_2 + \sigma_3)} - \tfrac{1}{2} g_0{}^2 e^{5/2 \, \sigma_1 + \sqrt{3}/2 \, \sigma_2} + \notag \\
  & - \tfrac{1}{2}  (g_6 + h_3 \chi)^2 e^{- \sigma_1/2 + 3 \sqrt{3} /2 \, \sigma_2} ] \,. \label{L2}
 \end{align}
Note that there are no topological Chern-Simons terms in the democratic formulation of IIA \cite{demo}; the kinetic terms for the different R-R potentials suffice.
These are therefore all the contributions from the ten-dimensional bulk action.

In addition to the bulk, one must also include the orientifold planes induced by \eqref{fields} and \eqref{coords}. We further allow for a number of D6-branes with the same orientation (ignoring the world-volume excitations). These give rise to the following contributions to the scalar potential:
 \begin{align}
  \hat S_{\rm O6/D6} = 2 \pi N \int d^7 x [ \, e^{3/4 \, \hat \phi} \sqrt{- \hat g_7} \, ] \quad \rightarrow \quad {\mathcal L}_3 = \sqrt{-g} \, [ \, N \, e^{3/4 \, \sigma_1 + 3 \sqrt3/4 \, \sigma_2 + \sqrt{3} /2 \, \sigma_3} ] \,, \label{L3}
 \end{align}
where $N = 2 N_{\rm O6} - N_{\rm D6}$. An orientifolded three-torus
has $2^3$ fixed points under \eqref{coords} and would lead to
$N_{\rm O6} = 8$. The unusual dilaton coupling stems from the fact
that we are using Einstein frame. Furthermore, we have not included
the Wess-Zumino term, as this will not contribute to the
four-dimensional action. The total resulting action consists of the
three pieces \eqref{L1}, \eqref{L2} and \eqref{L3}.

The Bianchi identities for the different field strengths read
 \begin{align}
  d \hat H = 0 \,, \quad d \hat G_{2n+2} = \hat H \wedge \hat G_{2n} \,.
 \end{align}
These are satisfied by the Ans\"{a}tze above modulo the following two points. The first is that, due to the Wess-Zumino term of the O6-planes and D6-branes that involves $\hat C_7$, the Bianchi identity of $\hat G_2$ is modified:
 \begin{align}
  d \hat G_{2} = \hat G_0 \, \hat H - N \, dx^7 \wedge dx^8 \wedge dx^9 \,, \qquad \Rightarrow \qquad
  g_0 h_3 = N \,, \label{tadpole}
 \end{align}
leading to a tadpole condition that will be essential. Furthermore,
the reader might worry about the Bianchi identity for the electric
part of $\hat G_6$, which does not vanish identically. However, this
will be proportional to the four-dimensional field equation for
$\chi$ and vanishes on-shell.

Turning to the three pieces of which the action consists, we can now
appreciate the beauty of the orientifold reduction and the
underlying supersymmetry. The contribution due to the orientifold is
such that the scalar potential terms \eqref{L2} and \eqref{L3}
involving $g_0$ and $h_3$ can be combined into a square. This
crucially relies on the tadpole condition \eqref{tadpole}. The
scalar potential is now a positive definite sum of two squares.

The orientifold breaks half of supersymmetry and the resulting
four-dimensional description is an $\mathcal N = 4$ supergravity.
Since our truncation to $SL(3) \times SL(3)$ singlets
coincides with that of the previous section, there must be a
relation to the model discussed there. 
Indeed the two can be related by the following field redefinition for the $\sigma$'s:
 \begin{align}
  \left( \begin{array}{c} \phi \\ \varphi_1 \\ \varphi_2 \end{array} \right)
   = \frac{1}{4 \sqrt{2}} \left( \begin{array}{ccc}
   \sqrt{2} & 3 \sqrt{3} & - \sqrt{3} \\
   \sqrt{6} & 1 & 5 \\
   -2 \sqrt{6} & 2 & 2 \end{array}\right)
  \left(  \begin{array}{c} \sigma_1 \\ \sigma_2 \\ \sigma_3 \end{array} \right) \,,
 \end{align}
in terms of the $SL(2)$ dilaton $\phi$ and the $SO(6,6)$ dilatons $\varphi_1$ and $\varphi_2$ of the previous section.  Furthermore, one must identify the gauge parameters of both models as
 \begin{align}
  ( f_{+ 123}, f_{- 456}, f_{+ 456}) = \sqrt{2} (g_0, h_3, g_6) \,. \label{gaugeparameters}
 \end{align}
These redefinitions turn the Lagrangian consisting of \eqref{kin2}
and \eqref{pot} into that consisting of \eqref{L1}, \eqref{L2} and
\eqref{L3}. Moreover, since the $SL(3) \times SL(3)$ invariant
model defines a unique $\mathcal N = 4$ gauged supergravity, this
connection extends to the full theory: an orientifold reduction that
retains all fields and includes these three fluxes will lead to an
$\mathcal N = 4$ supergravity with gauge parameters
\eqref{gaugeparameters}.

Our simple orientifold reduction therefore leads to an $\mathcal N =
4$ supergravity with $CSO(1,0,3) \times CSO(1,0,3)$ gauge group, 
where the two gauge factors have a non-vanishing
relative $SL(2)$ angle. Note that the tadpole condition implies a
relation on the gauge parameters: they have to be of the form
\eqref{gaugeparameters} with $g_0$, $h_3$, $g_6$ integer and subject
to $g_0 h_3 = N$. Furthermore, the condition on the signs of $f_{+
123}$ and $f_{- 456}$ of the
previous section justifies our assumption that $g_0 h_3$ is
positive.

\section{Discussion and outlook}

In the previous sections we have seen that the simple IIA orientifold
reduction with fluxes $(g_0, h_3, g_6)$ leads to the ${\mathcal N} = 4$ supergravity with 
$CSO(1,0,3) \times CSO(1,0,3)$ gauge group of \cite{CSO}. The
two gauge factors generically are at a non-vanishing $SL(2)$ angle with respect to each other, leading to moduli stabilisation. 
From the orientifold side, this important feature 
was achieved by a collaboration of
contributions to the scalar potential from the IIA bulk action \eqref{L2} and 
the local source terms \eqref{L3} due to the orientifolding. 
In order to avoid this, one must tune the
O6/D6 content such that $N = 0$, in which case the two gauge groups
have the same angle or one of the them disappears. Thus we have
clarified the higher-dimensional origin of the important $\mathcal N=4$ phenomenon
of $SL(2)$ angles. Our simple model demonstrates that such gaugings at angles will be a
generic outcome of IIA orientifold reductions.

Due to T-duality our results can be related to other orientifold
cases. For instance, consider the case where we T-dualise in the three toroidal
directions $(x^4, x^5, x^6)$ of the O6-plane worldvolume. The
resulting IIB reduction involves an O3-plane and has been studied at
length in e.g.~\cite{KST, *Frey, D'Auria1, *D'Auria2}. Our results have a clear
counterpart in this IIB case. The parameters $g_0$, $h_3$ and $g_6$
now come from the IIB three-form components $\hat G_{456}$, $\hat
H_{789}$ and $\hat G_{789}$, respectively. The tadpole condition
relates D3-branes and O3-planes to a contribution due to the complex
three-form flux, and the resulting action also contains a sum of
squares. In this case the vanishing of the squares corresponds to
the well-known imaginary self-duality condition on the three-form
flux. Again the non-trivial $SL(2)$ angles play an important role in
the stabilisation of moduli.

On the other hand, one could consider T-duality in any of the directions $(x^7,x^8,x^9)$ transverse to the O6-plane. In contrast to the previous case, T-duality in these directions does not leave the three-form flux invariant. Instead it has been argued that this will be transformed into geometric or even non-geometric flux \cite{Wecht}. Therefore T-duality in the transverse directions, giving rise O7-, O8- or O9-planes, does not lead to the simple reductions we considered with only gauge fluxes.

Coming back to the O6-plane, the reduction to four dimensions can in
fact be split up in two steps. The first consists of the reduction
over the transverse space of the orientifold, while the second
reduces over its three toroidal world-volume directions. One could
stop after the first step and thus obtain a seven-dimensional
half-maximal supergravity theory with two parameters $g_0$ and $h_3$
(the remaining $g_6$ only shows up after the second step). The
gaugings of this theory are encoded in representations $\xi_m$ and
$f_{mnp}$ of $SO(3,3)$, while there is a single topological mass
parameter $m$ \cite{Nutma}. Restricting to $SL(3,\mathbb{R})$
invariant components leads to $f_{123}$ and $f_{\bar 1 \bar 2 \bar
3}$ in addition to the mass parameter. The product of the two gauge
parameters vanishes due to the Jacobi identity. Therefore, the
orientifold parameters $g_0$ and $h_3$ are to be identified with $m$
and e.g.~$f_{123}$. A subsequent normal toroidal
reduction to four dimensions leads to the theory that we have
considered in this paper.

Two lessons can be drawn from this discussion. Firstly, the $D=7$ topological mass parameter $m$ has a higher-dimensional origin from orientifold reductions. Secondly, gaugings at angles in four dimensions are induced by a toroidal reduction of the seven-dimensional massive theory. To our knowledge, this is the first time that 
a higher-dimensional origin of gaugings at angles from $4 < D \leq 7$ half-maximal supergravity has been put forward. It would be interesting to investigate this connection in more detail. Due to the above discussion involving O$p$-planes with $p>6$ we do not expect such an origin from dimensions higher than seven. This ties in nicely with the absence of mass parameters in these theories \cite{Nutma}.

In this paper we have restricted ourselves to a very simple
truncation to $SL(3) \times SL(3)$ singlets. Needless to say this
can be relaxed to allow for many more possibilities
\cite{D'Auria1, *D'Auria2, AFT1,*AFT2,*DAF}: different components of gauge fluxes can be
turned on and one could reduce over twisted tori with non-vanishing
geometric fluxes $\omega$. This would lead to additional structure
constants, inducing different gaugings of the four-dimensional $\cN
= 4$ theory. For instance, including $\omega$ and $\hat G_2$ fluxes
in a specific way could lead to $CSO(3,0,1) \times CSO(3,0,1)$
gaugings \cite{DKPZ}. It would be interesting to investigate a
possible relation to the $SU(2) \times SU(2)$ reduction of
\cite{Lust}. Furthermore, such reductions might give a
higher-dimensional origin to the unstable De Sitter vacua of
\cite{Westra1, *Westra2}.

Finally, one can consider orientifold reductions that break more
supersymmetry. It would be of great interest if one could find
e.g.~a string-theoretic origin for the stable De Sitter vacua in
$\mathcal N = 2$ supergravity \cite{Fre, *Ogetbil}, for which
non-trivial $SL(2)$ angles are a necessary ingredient.

\section*{Acknowledgements}

We thank Giuseppe Dibitetto, Thomas van Riet and Mees de Roo for very stimulating discussions.
The work of DR is supported by a VIDI grant from the Netherlands Organisation for Scientific Research (NWO).

\providecommand{\href}[2]{#2}\begingroup\raggedright\endgroup


\begin{thebibliography}{10}

\bibitem{review}
 This has been reviewed in e.g.

\bibitem{Grana}
M.~Grana,  {\em {Flux compactifications in string theory: A comprehensive
  review}}, Phys. Rept. {\bf 423} (2006) 91--158
[\href{http://www.arXiv.org/abs/hep-th/0509003}{{\tt hep-th/0509003}}].

\bibitem{Douglas}
M.~R. Douglas and S.~Kachru,  {\em {Flux compactification}}, Rev. Mod. Phys.
  {\bf 79} (2007) 733--796
[\href{http://www.arXiv.org/abs/hep-th/0610102}{{\tt hep-th/0610102}}].

\bibitem{see}
 Early references are e.g.

\bibitem{KKLT}
S.~Kachru, R.~Kallosh, A.~Linde and S.~P. Trivedi,  {\em {De Sitter vacua in
  string theory}}, Phys. Rev. {\bf D68} (2003) 046005
[\href{http://www.arXiv.org/abs/hep-th/0301240}{{\tt hep-th/0301240}}].

\bibitem{Burgess}
C.~P. Burgess, R.~Kallosh and F.~Quevedo,  {\em {de Sitter String Vacua from
  Supersymmetric D-terms}}, JHEP {\bf 10} (2003) 056
[\href{http://www.arXiv.org/abs/hep-th/0309187}{{\tt hep-th/0309187}}].

\bibitem{Kallosh}
R.~Kallosh, A.~D. Linde, S.~Prokushkin and M.~Shmakova,  {\em {Gauged
  supergravities, de Sitter space and cosmology}}, Phys. Rev. {\bf D65} (2002)
  105016
[\href{http://www.arXiv.org/abs/hep-th/0110089}{{\tt hep-th/0110089}}].

\bibitem{Westra1}
M.~de~Roo, D.~B. Westra and S.~Panda,  {\em {De Sitter solutions in N = 4
  matter coupled supergravity}}, JHEP {\bf 02} (2003) 003
[\href{http://www.arXiv.org/abs/hep-th/0212216}{{\tt hep-th/0212216}}].

\bibitem{Westra2}
M.~de~Roo, D.~B. Westra, S.~Panda and M.~Trigiante,  {\em {Potential and
  mass-matrix in gauged N = 4 supergravity}}, JHEP {\bf 11} (2003) 022
[\href{http://www.arXiv.org/abs/hep-th/0310187}{{\tt hep-th/0310187}}].

\bibitem{Fre}
P.~Fre, M.~Trigiante and A.~Van~Proeyen,  {\em {Stable de Sitter vacua from N =
  2 supergravity}}, Class. Quant. Grav. {\bf 19} (2002) 4167--4194
[\href{http://www.arXiv.org/abs/hep-th/0205119}{{\tt hep-th/0205119}}].

\bibitem{Ogetbil}
O.~Ogetbil,  {\em {Stable de Sitter Vacua in 4 Dimensional Supergravity
  Originating from 5 Dimensions}}, Phys. Rev. {\bf D78} (2008) 105001
[\href{http://www.arXiv.org/abs/0809.0544}{{\tt 0809.0544}}].

\bibitem{dRW}
M.~de~Roo and P.~Wagemans,  {\em {Gauge matter coupling in N=4 supergravity}},
  Nucl. Phys. {\bf B262} (1985)
644.

\bibitem{D'Auria1}
R.~D'Auria, S.~Ferrara and S.~Vaula,  {\em {N = 4 gauged supergravity and a IIB
  orientifold with fluxes}}, New J. Phys. {\bf 4} (2002) 71
[\href{http://www.arXiv.org/abs/hep-th/0206241}{{\tt hep-th/0206241}}].

\bibitem{D'Auria2}
R.~D'Auria, S.~Ferrara, F.~Gargiulo, M.~Trigiante and S.~Vaula,  {\em {N = 4
  supergravity Lagrangian for type IIB on T**6/Z(2) in presence of fluxes and
  D3-branes}}, JHEP {\bf 06} (2003) 045
[\href{http://www.arXiv.org/abs/hep-th/0303049}{{\tt hep-th/0303049}}].

\bibitem{AFT1}
C.~Angelantonj, S.~Ferrara and M.~Trigiante,  {\em {New D = 4 gauged
  supergravities from N = 4 orientifolds with fluxes}}, JHEP {\bf 10} (2003)
  015
[\href{http://www.arXiv.org/abs/hep-th/0306185}{{\tt hep-th/0306185}}].

\bibitem{AFT2}
C.~Angelantonj, S.~Ferrara and M.~Trigiante,  {\em {Unusual gauged
  supergravities from type IIA and type IIB orientifolds}}, Phys. Lett. {\bf
  B582} (2004) 263--269
[\href{http://www.arXiv.org/abs/hep-th/0310136}{{\tt hep-th/0310136}}].

\bibitem{DAF}
G.~Dall'Agata and S.~Ferrara,  {\em {Gauged supergravity algebras from twisted
  tori compactifications with fluxes}}, Nucl. Phys. {\bf B717} (2005) 223--245
[\href{http://www.arXiv.org/abs/hep-th/0502066}{{\tt hep-th/0502066}}].

\bibitem{SW}
J.~Schon and M.~Weidner,  {\em {Gauged N = 4 supergravities}}, JHEP {\bf 05}
  (2006) 034
[\href{http://www.arXiv.org/abs/hep-th/0602024}{{\tt hep-th/0602024}}].

\bibitem{Prezas}
J.-P. Derendinger, P.~M. Petropoulos and N.~Prezas,  {\em {Axionic symmetry
  gaugings in N = 4 supergravities and their higher-dimensional origin}}, Nucl.
  Phys. {\bf B785} (2007) 115--134
[\href{http://www.arXiv.org/abs/0705.0008}{{\tt 0705.0008}}].

\bibitem{DKPZ}
J.-P. Derendinger, C.~Kounnas, P.~M. Petropoulos and F.~Zwirner,  {\em
  {Superpotentials in IIA compactifications with general fluxes}}, Nucl. Phys.
  {\bf B715} (2005) 211--233
[\href{http://www.arXiv.org/abs/hep-th/0411276}{{\tt hep-th/0411276}}].

\bibitem{VZ}
G.~Villadoro and F.~Zwirner,  {\em {N = 1 effective potential from dual
  type-IIA D6/O6 orientifolds with general fluxes}}, JHEP {\bf 06} (2005) 047
[\href{http://www.arXiv.org/abs/hep-th/0503169}{{\tt hep-th/0503169}}].

\bibitem{Thomas}
S.~S. Haque, G.~Shiu, B.~Underwood and T.~Van~Riet,  {\em {Minimal simple de
  Sitter solutions}}, Phys.\ Rev.\  D {\bf 79} (2009) 086005
[\href{http://www.arXiv.org/abs/0810.5328}{{\tt 0810.5328}}].

\bibitem{ET1}
H.~Nicolai and H.~Samtleben,  {\em {Compact and noncompact gauged maximal
  supergravities in three dimensions}}, JHEP {\bf 04} (2001) 022
[\href{http://www.arXiv.org/abs/hep-th/0103032}{{\tt hep-th/0103032}}].

\bibitem{ET2}
B.~de~Wit, H.~Samtleben and M.~Trigiante,  {\em {On Lagrangians and gaugings of
  maximal supergravities}}, Nucl. Phys. {\bf B655} (2003) 93--126
[\href{http://www.arXiv.org/abs/hep-th/0212239}{{\tt hep-th/0212239}}].

\bibitem{CSO}
M.~de~Roo, D.~B. Westra and S.~Panda,  {\em {Gauging CSO groups in N = 4
  supergravity}}, JHEP {\bf 09} (2006) 011
[\href{http://www.arXiv.org/abs/hep-th/0606282}{{\tt hep-th/0606282}}].

\bibitem{DeWolfe}
O.~DeWolfe, A.~Giryavets, S.~Kachru and W.~Taylor,  {\em {Type IIA moduli
  stabilization}}, JHEP {\bf 07} (2005) 066
[\href{http://www.arXiv.org/abs/hep-th/0505160}{{\tt hep-th/0505160}}].

\bibitem{Camara}
P.~G. Camara, A.~Font and L.~E. Ibanez,  {\em {Fluxes, moduli fixing and
  MSSM-like vacua in a simple IIA orientifold}}, JHEP {\bf 09} (2005) 013
[\href{http://www.arXiv.org/abs/hep-th/0506066}{{\tt hep-th/0506066}}].

\bibitem{Polchinski}
J.~Polchinski,  {\em {String theory. Vol. 2: Superstring theory and beyond}},
  Cambridge, UK: Univ. Pr. (1998) 531 p.

\bibitem{demo}
E.~Bergshoeff, R.~Kallosh, T.~Ortin, D.~Roest and A.~Van~Proeyen,  {\em {New
  Formulations of D=10 Supersymmetry and D8-O8 Domain Walls}}, Class. Quant.
  Grav. {\bf 18} (2001) 3359--3382
[\href{http://www.arXiv.org/abs/hep-th/0103233}{{\tt hep-th/0103233}}].

\bibitem{KST}
S.~Kachru, M.~B. Schulz and S.~Trivedi,  {\em {Moduli stabilization from fluxes
  in a simple IIB orientifold}}, JHEP {\bf 10} (2003) 007
[\href{http://www.arXiv.org/abs/hep-th/0201028}{{\tt hep-th/0201028}}].

\bibitem{Frey}
A.~R. Frey and J.~Polchinski,  {\em {N = 3 warped compactifications}}, Phys.
  Rev. {\bf D65} (2002) 126009
[\href{http://www.arXiv.org/abs/hep-th/0201029}{{\tt hep-th/0201029}}].

\bibitem{Wecht}
J.~Shelton, W.~Taylor and B.~Wecht,  {\em {Nongeometric Flux
  Compactifications}}, JHEP {\bf 10} (2005) 085
[\href{http://www.arXiv.org/abs/hep-th/0508133}{{\tt hep-th/0508133}}].

\bibitem{Nutma}
E.~A. Bergshoeff, J.~Gomis, T.~A. Nutma and D.~Roest,  {\em {Kac-Moody Spectrum
  of (Half-)Maximal Supergravities}}, JHEP {\bf 02} (2008) 069
[\href{http://www.arXiv.org/abs/0711.2035}{{\tt 0711.2035}}].

\bibitem{Lust}
C.~Caviezel {\em et al.},  {\em {On the Cosmology of Type IIA Compactifications
  on SU(3)- structure Manifolds}}, JHEP {\bf 04} (2009) 010
[\href{http://www.arXiv.org/abs/0812.3551}{{\tt 0812.3551}}].

\end{thebibliography}
\end{document}